\documentclass[a4paper,10pt]{article}
\usepackage{graphicx}

\begin{document}


\begin{titlepage}

\begin{center}
\Large{\bf{
Metastable states in a class of long-range Hamiltonian systems
}}
\end{center}

\vspace{5mm}
\begin{center}
\large{
Alessandro Campa$^{a}$, Andrea Giansanti$^{b\S}$,
Daniele Moroni$^{c}$
}
\end{center}
\vspace{5mm}
\begin{center}
\normalsize{
$^a$Laboratorio di Fisica, Istituto Superiore di Sanit\`a\\ and
 INFN Sezione di Roma1, Gruppo Collegato Sanit\`a\\Viale Regina
 Elena 299, 00161 Roma, Italy
}
\end{center}
\begin{center}
\normalsize{
$^b$Dipartimento di Fisica, Universit\`a di
 Roma ``La Sapienza''\\ and INFM Unit\`a di Roma1, \\
 Piazzale Aldo Moro 2, 00185 Roma, Italy
}
\end{center}
\begin{center}
\normalsize{
$^c$Department of Chemical Engineering,
 Universiteit van Amsterdam\\
 Nieuwe Achtergracht 166, 1018 WV Amsterdam, The Netherlands
}
\end{center}

\vspace{5mm}
\begin{center}
\large{\bf {Abstract}}
\end{center}
\vspace{2mm}
We numerically show that metastable states, similar to the 
\emph{Quasi Stationary States} found in the so called Hamiltonian Mean Field
Model, are also present in a generalized model in which $N$ classical spins (rotators)
interact through ferromagnetic couplings decaying as $r^{-\alpha}$, where $r$ is
their distance over a regular lattice. Scaling laws with $N$ are briefly discussed.
\vspace{5mm}

\noindent{\it {\bf keywords}: Hamiltonian dynamics; Long-range interactions;
Metastable states; Nonextensive statistical mechanics.}

\vspace{5mm}
\noindent$^{\S}$\small{Corresponding author: Andrea.Giansanti@roma1.infn.it}\\

\end{titlepage}

\normalsize
\setlength{\baselineskip}{20pt}


\section{Introduction}

The study of the dynamical properties of systems with long-range interactions
is important from a foundational point of view. It is well known that if the
potential energy of a classical hamiltonian system does not obey the so called
\emph{temperedness} and \emph{stability} conditions then the existence of a
statistical mechanics is questionable \cite{gal}. If the potentials are well
behaved it can be shown that the statistical thermodynamics of the system
is \emph{extensive} (i. e. the free energy density does not depend on the
number of degrees of freedom) and \emph{additive} (i. e. the free energy of
the system as a whole can be expressed as the sum of the free energies of
its component macroscopic parts) \cite{note1}.

The good potentials are repulsive at short distances and decay sufficiently
fast at large distances. Slowly decaying interactions may induce into the
system spatial and temporal correlations leading to anomalous relaxations
and to equilibrium distribution functions different from that of standard
statistical mechanics. The nonextensive generalization of Boltzmann-Gibbs
statistics, proposed by C. Tsallis \cite{T1,T2}, is a powerful parametrization
of the statistical mechanics of out-of-equilibrium states in open systems.
The formulas in this approach are labelled by the so called \emph{entropic
index} $q$, and in the limit $q=1$ the Boltzmann-Gibbs expressions are
recovered.

It has been also conjectured that this formalism could be applied to anomalous
states of conservative hamiltonian systems with long-range interactions
\cite{T2}. The so called quasi stationary states (QSS) of the Hamiltonian Mean
Field model (HMF), originally introduced by Ruffo and Antoni \cite{AR}, have
been recently investigated and discussed in this perspective by Latora,
Rapisarda and Tsallis \cite{LRT1,LRT2}. If particular initial conditions are
chosen for the model, microcanonical simulations reveal anomalous properties
\cite{ART}. For example anomalous diffusion is observed \cite{LR,LRR1,LRR2},
single-particle velocity distribution function are not gaussian
\cite{LRT1,LRT2,LR,LRR3} and the temperature as a function of time is locked
at a value which is dependent on $N$, and which is different from the canonical
equilibrium one \cite{LRT1,LRT2,LR,LRR2,LR2}, established analitically
\cite{AR}.
After a characteristic lifetime, those non-usual properties relax toward the
ones predicted by standard statistical mechanics. This relaxation has been
appealing interpreted as the irreversible passage from a non equilibrium state,
appropriately described by Tsallis' statistics, toward the equilibrium
Boltzmann-Gibbs state \cite{LRT1}.
Since the lifetime of the QSS has been shown to diverge with $N$, then in the
thermodynamic limit the system is frozen forever in a state characterized by
a $q$ different from 1.

In this paper we give numerical evidence that the stationary states are
also present in a system that generalizes the HMF model. In the system we
consider classical spins interact with couplings that decay as $r^{-\alpha}$,
where $r$ is the distance between two spins. If $\alpha \geq d$, the
euclidean dimension of the embedding space, then the interaction
is short-ranged, if $0 \leq \alpha < d$ it is long-ranged.
The spins are fixed at the nodes of a $d$-dimensional lattice; this avoids the
singularity of the potential terms at very short distances.

As we shall point out in sect. \ref{sec.model}, the system can be made
extensive through an appropriate rescaling function. Nevertheless, in the
long-range case, additivity might not hold, and anomalous dynamical effects
similar to those observed in the infinite range HMF model are expected.

The present study is the starting point of a project devoted to the
investigation of the effects of the range of the interactions, as controlled
by the exponent $\alpha$, on the dynamical properties of a class of systems for
which the HMF model is the infinite range limit.


\section{The model}\label{sec.model}

We consider the following classical hamiltonian
\begin{equation}\label{num2.1}
 H=\frac{1}{2} \sum_{i=1}^{N} L_i^2 \, + \, \frac{1}{2\tilde{N}}
 \sum_{i\neq j}^{N} \frac {1-\cos(\theta_i-\theta_j)}
 {r_{ij}^\alpha}
\end{equation}
of planar rotators (XY spins) on a generic $d$-dimensional lattice. At each
lattice site $i$ conjugate canonical coordinates $(L_i,\theta_i)$ represent
the angular momentum and the angular position of a rotator. Each of them has unit
moment of inertia and moves on a reference plane passing through its lattice
site. Parameter $\alpha\geq 0$ tunes the interaction and periodic boundary
conditions are enforced in the form of nearest image convention. For $\alpha=0$
the hamiltonian of the HMF model is recovered. The rescaling
parameter $\tilde{N}$ is a function of $\alpha,N,d$ and the lattice, and is
defined as $\tilde{N}=\sum_{j\neq i} 1/r_{ij}^\alpha$. Because of periodicity
the previous sum does not depend on the choice of the origin $i$.
The presence of this rescaling is necessary to make the system extensive, i.e.
to guarantee the existence of a bounded energy density in the thermodynamic
limit \cite{CGM}.

The model, that we tend to call the $\alpha-XY$ model, has been introduced by
Anteneodo and Tsallis \cite{AT}; its thermodynamics has been studied
computationally for $d=1$ \cite{TA} and its canonical partition function has
been computed analytically \cite{CGM,GMC} in the case $\alpha<d$. These last 
studies have shown that, through a rescaling function $\tilde{N}$,
 the canonical statistical mechanics of the $\alpha-XY$ can be mapped 
on that of the HMF model.
In particular, both models have a second-order phase transition
between a supercritical disordered phase and a low energy ferromagnetic
ordered phase, with the same critical energy $U_c=3/4$.

The study of the chaotic properties of the supercritical $\alpha-XY$ model has
remarkably shown a universal reduction of mixing in the thermodynamic limit,
controlled by the parameter $\alpha/d$ \cite{AT,GMC,CGMT}.
These results have found a theoretical setting in a paper by
Firpo and Ruffo \cite{FR}.


\section{Simulation details and results}

We have decided to carry on the search for the existence of metastable states
in the generalized model (\ref{num2.1}) along the same lines of the studies
conducted on the HMF model. A previous brief inspection of the single-particle
velocity distribution functions had revealed a non-Gaussian distribution, also for $\alpha
\neq 0$ \cite{GMC}. Here we focus on the behaviour of the instantaneous
temperature T(t) defined as both a time and ensemble average:
\begin{equation}\label{num3.1}
T(t)=\langle \int_{t_0}^t dt' 2K(t')/N \rangle,
\end{equation}
where $K$ is the total kinetic energy, $t_0$ is an equilibration time, and the
external average $\langle\cdot\rangle$ is over different initial conditions, all of the ''water-bag'' form:
$\theta_i=0$ for all $i$ and $L_i$ uniformly distributed. We have simulated the
system in dimension $d=1$ for different $\alpha$ and $N$. The simulations are performed
at fixed energy integrating the equations of motion through a $4^{th}$ order
symplectic algorithm \cite{Yo} with time-step $dt=0.02$ which gives an energy
conservation $\Delta E/E\sim10^{-7}$. We chose $t_0=100$ and rescaled the
velocities to fix the energy density value at $U=0.69$. For short, we have followed
 the protocol that was successful in detecting metastable states
in earlier works \cite{LRT1,LRT2,ART,LR,LRR1,LRR2,LRR3,LR2,LRR4}.

We have first run a few simulations for one fixed $\alpha=0.4$, increasing
$N$ up to $N=2000$. The results are shown in Fig. \ref{fig1}. They indeed
confirm the existence of a temperature plateau which, after a given time, relaxes
towards the canonical temperature (known to be at $T=0.476$). 
The length of the plateau
increases for increasing $N$ and characterises therefore the lifetime $\tau$
of a true metastable (or "Quasi stationary" (QSS)) state. 
The height of plateau gives the ''metaequilibrium''
temperature $T_{QSS}$ and is found to decrease with increasing $N$ as in
the HMF model \cite{LRT2}. However how these two quantities
$\tau$ and $T_{QSS}$ scale with $N$ is not clear and still 
under investigation.

We have to remark that the computation of the quantity (\ref{num3.1}) is
strongly affected by fluctuations. Both averages (time, ensemble) in formula
(\ref{num3.1}) appear necessary to have a good statistics and thus a smooth
curve. We have checked that averaging over a set of $100$ initial conditions
led to a definite smooth $T$  vs  $t$ curves. Averaging over an increasing
number of initial conditions has the effect of slightly shifting the curves,
but $T_{QSS}$ and $\tau$ change within a few percent.  
For 100 averages
the curve appears definite but only for averages over 1000 initial conditions
the height of the plateau (but not the length) seems not to be affected any
more. 
We expect that at large $N$ one has to average over less initial conditions, 
but still the task is computationally expensive
even if our integration of the forces is of order $N\ln N$, due to the use
of Fast Fourier Transforms.

We decided then to fix the number of rotators at $N=500$ and concentrate our
computational resources to the study of the effects of changes in $\alpha$. 
We simulated ten
values of $\alpha$ between 0 and 1 and averaged over 100 initial conditions.
We report the resulting temperature curves in Fig. \ref{fig2}.

We observe slight discrepancies between our curve for $\alpha=0$ and the
one found in \cite{LRT1,LRT2} because the height of our plateau looks
a bit higher. We think the difference is due to a poor average
on initial conditions which as said before can affect this quantity.

From the curves in Fig. \ref{fig2} we computed the lifetime $\tau$ of the
metastable state for the various $\alpha$'s. We defined $\tau$ to be the point
where a sharp increase in the curve is found. We point out that this
definition is purely qualitative. Smoother data from a better (and harder)
sampling should lead to a more precise definition and thus a better
estimate of $\tau$. We report the results in Fig. \ref{fig3}. Note that $\tau$
at $\alpha=0$ is around 1000, a value consistent with the HMF results 
(see fig. 1b
of ref. (\cite{LRT2}) and fig. 2 of ref.(\cite{LRT1})). As soon as $\alpha$
is increased, reducing the range of the interaction, 
an abrupt increase of $\tau$ is observed. From $\alpha=0.1$ on $\tau$ decreases
almost exponentially. It would be interesting to carefully investigate the effect
of small $\alpha$s in a more extended study.


\section{Discussion}

\begin{figure}[hb]
\begin{center}
\includegraphics[width=\textwidth,keepaspectratio]{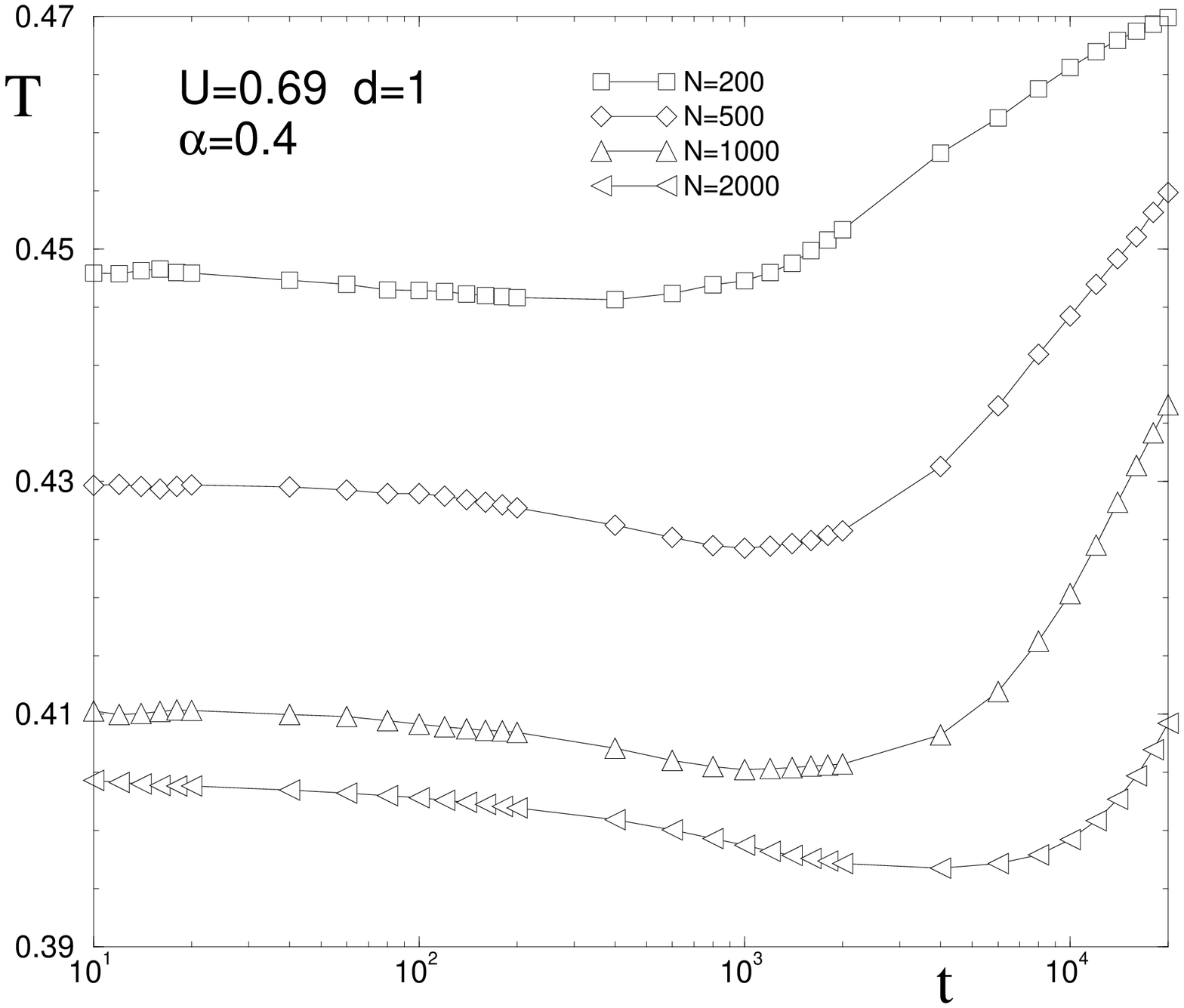}
\end{center}
\caption{$T$ vs $t$ for various $N$ and $\alpha=0.4$. Lifetime of the
metastable state increases with $N$ also for $\alpha \neq 0$. Plateau height
also decreases in accordance with $\alpha=0$ results. The $N=200,500,1000,2000$
curves are averaged respectively over $300,100,100,50$ initial conditions.}
\label{fig1}
\end{figure}

In this paper we have shown that metastable states are also present in the
$\alpha-XY$ model; this was quite expected. Qualitatively, the quasi stationary
states behave similarly to those observed in the HMF model; in particular, from
Fig. \ref{fig1} we see that lifetimes tend to increase with $N$ and the
\emph{plateau} temperatures tend to decrease with $N$; we have checked with 
short runs up to $N=100000$ and $\alpha=0.9$ that the plateau tend to the 
same value of $T_{\infty}=0.38$, that
has been obtained extending the upper-critical universal caloric curve 
of the HMF
model (corresponding to $\alpha=0$) into the lower-critical region \cite{LRT2}.
From Fig. \ref{fig3} we can say that the lifetime of the QSS tends to become
very small as $\alpha$ tends to $1$, and that confirms that the existence
and stabilization of these states is due to the range of the interactions.

\begin{figure}[hb]
\begin{center}
\includegraphics[width=\textwidth,keepaspectratio]{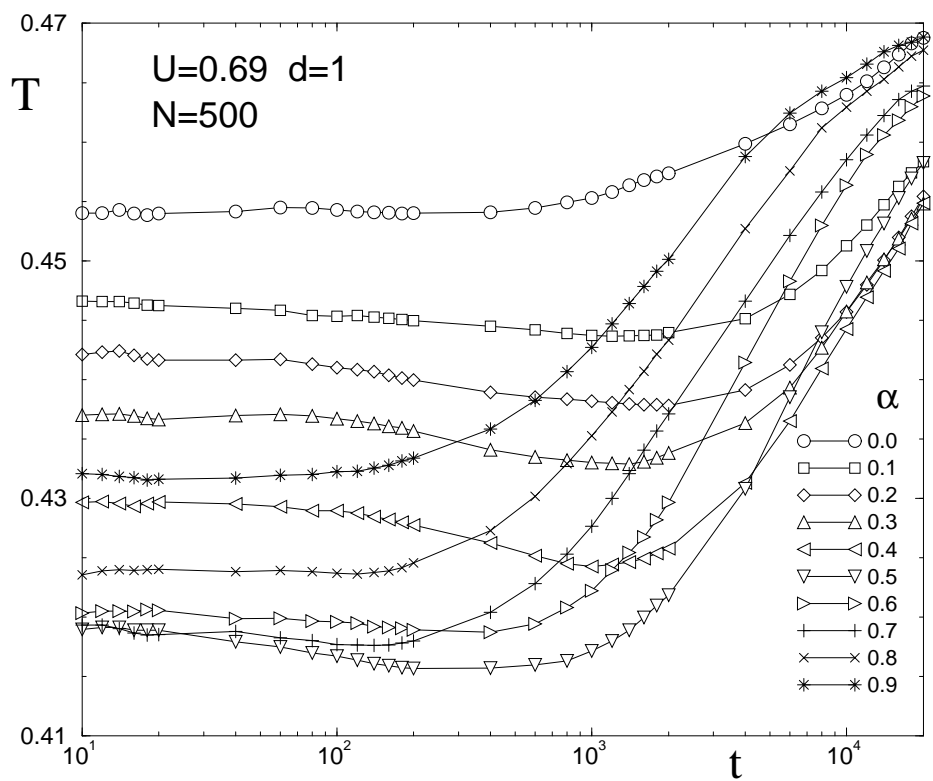}
\end{center}
\caption{$T$ vs $t$ for various $\alpha$ and $N=500$. Lifetimes of the
metastable states and plateau height depend on $\alpha$, the latter also
non-monotonically. All curves are averaged over 100 initial conditions.}
\label{fig2}
\end{figure}

\begin{figure}[hb]
\begin{center}
\includegraphics[width=\textwidth,keepaspectratio]{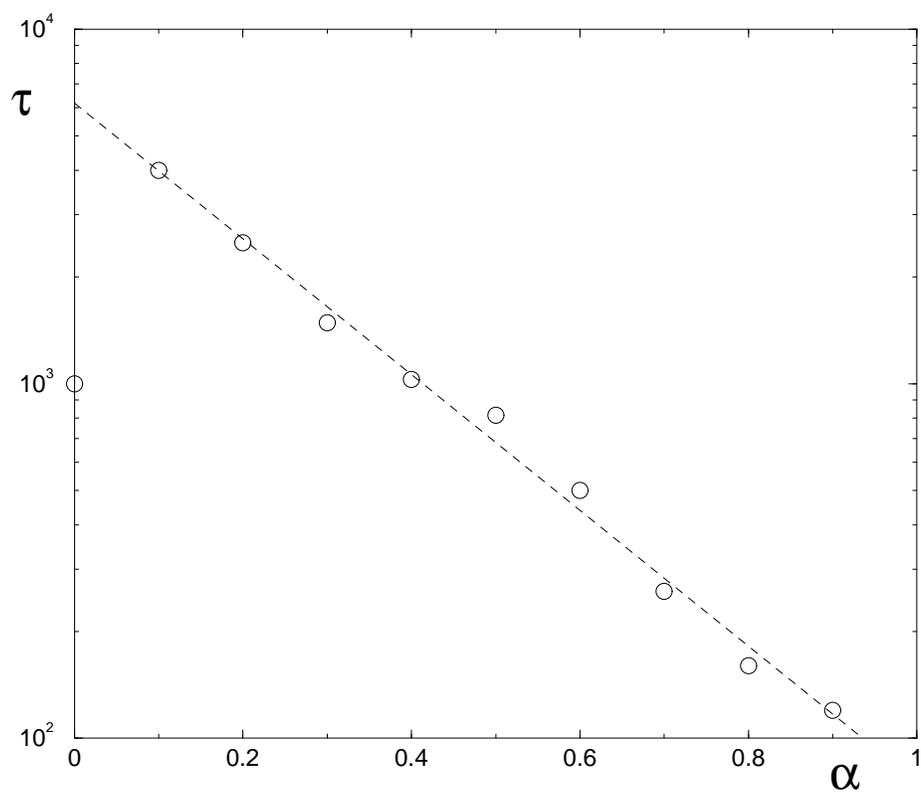}
\end{center}
\caption{QSS lifetime $\tau$ vs $\alpha$ as extracted from figure \ref{fig2}.
The dashed line is an exponential fit.}
\label{fig3}
\end{figure}

We have made efforts to rescale the data in Fig. \ref{fig1} and Fig.
\ref{fig2}, using the scaling factor $\tilde{N}$. Our aim was to show that the
scaling laws observed in the HMF models are also valid in the $\alpha-XY$
model,  but the simple ways we have attempted gave no result. In fact, the exponential
fit in Fig. \ref{fig3} shows that the lengths of the temperature plateau
do not scale with $\tilde{N}$ (that is with $N^{1-\alpha}$\cite {TA}). On the other hand
we have to say that $\tilde{N}$ is a rescaling function that was successfully
 applied, as an ''effective number of interacting degrees of freedom'' \cite{GMC},
 to the computation of 
equilibrium canonical quantities. There is no reason that scaling by the same function
should reveal some universality of the non-equilibrium dynamical features in this
class of models. More work is needed to search for such a universal behaviour.

We gratefully acknowledge warm, friendly and illuminating discussions
with Vito Latora, Andrea Rapisarda, Stefano Ruffo and Constantino Tsallis.



\clearpage

%


\end{document}